\documentclass[12pt]{article}
\pdfoutput=1
\usepackage{subfigure}
\usepackage{amssymb,amsmath}
\usepackage{graphicx}
\usepackage{color}
\usepackage[colorlinks=true
,urlcolor=blue
,citecolor=blue
,linkcolor=blue
,pagecolor=blue
,linktocpage=true
,pdfproducer=medialab
]{hyperref}
\usepackage[a4paper,width=15.2cm]{geometry}

\makeatletter \renewcommand{\@dotsep}{10000} \makeatother

\newcommand{\beq}{\begin{equation}}
\newcommand{\eeq}{\end{equation}}
\newcommand{\bea}{\begin{eqnarray}}
\newcommand{\eea}{\end{eqnarray}}

\newcommand\npb[3]{{\it Nucl.\ Phys.\ }{\bf B #1} (#2) #3}
\newcommand\plb[3]{{\it Phys.\ Lett.\ }{\bf B #1} (#2) #3}

\newcommand\jhep[3]{{\it J. High Energy Phys.\ }{\bf #1} (#2) #3}

\begin{document}

\begin{center}

 {\Large\bf  GUT-Inspired Supersymmetric Model \\ for   $h\rightarrow \gamma \gamma$ and   Muon $g-2$
 } \vspace{1cm}

{\large   M. Adeel Ajaib$^{a,}$\footnote{ E-mail: adeel@udel.edu}, Ilia Gogoladze$^{b,}$\footnote{E-mail: ilia@bartol.udel.edu\\
\hspace*{0.5cm} On  leave of absence from: Andronikashvili Institute
of Physics, 0177 Tbilisi, Georgia.}, and   Qaisar Shafi$^{b,}$\footnote{ E-mail:
shafi@bartol.udel.edu} } \vspace{.5cm}

{\baselineskip 20pt \it
$^a$Department of Physics and Astronomy, Ursinus College, Collegeville, PA 19426\\
$^b$Bartol Research Institute, Department of Physics and Astronomy, \\
University of Delaware, Newark, DE 19716, USA  } \vspace{.5cm}

\vspace{1.5cm}

 {\bf Abstract}
\end{center}

 We study a GUT-inspired {supersymmetric} model with non-universal gaugino masses that can explain the observed {muon }$g-2$ anomaly while simultaneously accommodating an enhancement {or} suppression in the $h\rightarrow\gamma\gamma$ {decay} channel. In order to accommodate these observations {and $m_h \simeq 125-126$ GeV}, the model {requires} a spectrum consisting of {relatively} light sleptons whereas the colored sparticles are heavy.
  The predicted stau mass range  corresponding to $R_{\gamma \gamma}\ge 1.1$ is $100 {\rm \ GeV} \lesssim m_{\tilde{\tau}} \lesssim 200 {\rm \ GeV}$. { The constraint on the slepton masses, particularly on the smuons, arising from considerations of muon $g-2$ is somewhat milder. The slepton masses in this case are predicted to lie in the few hundred GeV range.} The colored sparticles  {turn out to be considerably} heavier with $ m_{\tilde{g}} \gtrsim 4.5 {\rm \ TeV} $ and $ m_{\tilde{t}_1} \gtrsim 3.5 {\rm \ TeV} $, which makes it challenging for these to be observed at {the}  14 TeV LHC.

\newpage

\renewcommand{\thefootnote}{\arabic{footnote}}
\setcounter{footnote}{0}



\section{\label{ch:introduction}Introduction}

{The}  ATLAS and CMS {experiments at the LHC} have independently reported the discovery \cite{:2012gk, :2012gu} of a  Standard Model (SM)--like Higgs boson  of mass $m_h \simeq 125-126$ GeV
 using the combined 7~TeV and 8~TeV data. This discovery is compatible with low ({TeV}) scale supersymmetry \cite{Martin:1997ns}.  At the same time,
 after the first LHC {run} we have the following lower bounds on the {gluino and squark} masses~\cite{Aad:2012fqa,Chatrchyan:2012jx}
\begin{equation}
  m_{\tilde{g}} \gtrsim  1.4~{\rm TeV}~ ({\rm for}~ m_{\tilde{g}}\sim m_{\tilde{q}})~~~ {\rm and}~~~
m_{\tilde{g}}\gtrsim 0.9~{\rm TeV}~ ({\rm for}~ m_{\tilde{g}}\ll
m_{\tilde{q}}).
\label{bound1}
\end{equation}
{In some well motivated SUSY models the gluino is {the} NLSP in which case $m_{\tilde g} \gtrsim 400$ GeV \cite{Gogoladze:2009bn}. }
These bounds combined with the bound of 125 GeV on the lightest CP even Higgs boson mass place stringent constraints on the slepton  and gaugino (bino or wino) {mass} spectrum in {several} well studied scenarios such as constrained MSSM (cMSSM) \cite{Kane:1993td}, NUHM1 \cite{Baer:2004fu} and  NUHM2 \cite{nuhm2}. {In particular}, as we shall show later, in the above mentioned  models, the first two generation sleptons are predicted to be more than  1 TeV in order to accommodate the light CP even Higgs with 125 GeV mass. The stau leptons can still be relatively light due to {a relatively large} trilinear soft supersymmetry breaking (SSB) A-term.

 There are several motivations to study models that allow for the sleptons be as light as $\sim $ 100 GeV. For instance, the SM prediction for the anomalous magnetic moment of the muon, $a_{\mu}=(g-2)_{\mu}/2$ (muon $g-2$) \cite{Davier:2010nc}, {shows} a discrepancy with the experimental results \cite{Bennett:2006fi}:
\begin{eqnarray}
\Delta a_{\mu}\equiv a_{\mu}({\rm exp})-a_{\mu}({\rm SM})= (28.6 \pm 8.0) \times 10^{-10}.
\end{eqnarray}
If supersymmetry is to offer a  solution to this discrepancy, the   smuon and gaugino (bino or wino) SSB masses  should  be ${\cal O}(100)$ GeV or so \cite{Moroi:1995yh}.
 Thus, it is hard to simultaneously  explain the  observed Higgs boson mass and resolve the muon $g-2$ anomaly if  we consider CMSSM, NUHM1 or NUHM2, since in all these cases, the slepton masses are {larger} than 1 TeV.

Recently, there have been several attempts  to reconcile this {apparent} tension between muon $g-2$ and the Higgs boson mass within the  MSSM framework by assuming {non-universal SSB mass terms for the gauginos \cite{Gogoladze:2014cha,Akula:2013ioa} or the sfermions \cite{Babu:2014sga,Ibe:2013oha} at the GUT scale}. {Indeed, a} simultaneous explanation of $m_h$ and muon $g-2$ is possible \cite{Ajaib:2014ana} {in the presence of} $t-b-\tau$ Yukawa coupling unification condition  \cite{yukawaUn}. It has been shown \cite{Baer:2004xx} that constraints from FCNC processes are very mild and easily satisfied   for the case {in which}  the third generation sfermion masses  are split from those of the first two generations. However, {if } the muon $g-2$ anomaly and the Higgs boson mass are simultaneously explained with non-universal gaugino and/or sfermion masses, the correct relic abundance of neutralino dark matter is typically not obtained \cite{Ibe:2013oha}. Consistency with the observed dark matter abundance would further constrain the SUSY parameter space.

The Higgs decay channel $h\rightarrow \gamma\gamma$ {in recent times} attracted a {fair }amount of attention \cite{Carena:2012xa} because of the {apparent} deviation compared to the SM {prediction}. Currently,  the deviation from the SM prediction has significantly reduced but has  not completely disappeared.  For example, the ATLAS collaboration reported $\mu_{\gamma\gamma}
= 1.17\pm 0.27$ \cite{ATLAS-mugg-1}, where $\mu_{\gamma\gamma}=\frac{\sigma
(pp\rightarrow h\rightarrow \gamma\gamma)}{\sigma (pp\rightarrow h\rightarrow
\gamma\gamma)^{SM}}$. The  CMS collaboration reported a
best-fit signal strength in their main analysis $\mu_{\gamma\gamma} = 1.14^{+ 0.26}_{- 0.23}$ \cite{CMS-mugg-2}.
 On the other hand, a cut-based analysis
by CMS produced $\mu_{\gamma\gamma}=1.29^{+ 0.29}_{- 0.26}$, which is a slightly different value. This enhancement or suppression in the $h\rightarrow\gamma\gamma$ channel with respect to the SM  {may provide a clue} for physics beyond the SM if it is confirmed in the second LHC run. It is known that in order to accommodate  an enhancement or suppression in the  $h\rightarrow \gamma\gamma$ decay channel in the framework of MSSM, the stau is the one of the best candidates, and {its mass has to be} around 200 GeV or so. It is problematic to accommodate an enhancement or suppression in the $h\rightarrow \gamma \gamma $ decay channel in the framework of CMSSM, NUHM1 or NUHM2 models. { In this paper we present a GUT inspired model which explains the observed $g-2$ anomaly while simultaneously accommodating an enhancement or suppression in the $h\rightarrow\gamma\gamma$ channel. }\

The paper is organized as follows: In Section \ref{pheno} we describe the phenomenological constraints and the scanning procedure we implement in our analysis. In Section \ref{slepton} we {provide }motivations for the model used in this paper by briefly reviewing the status of the muon $g-2$ anomaly and $h\rightarrow\gamma\gamma$ in CMSSM and NUHM2. Our results for the $h\rightarrow\gamma\gamma$ channel in {the proposed} model are presented in Section \ref{Rgg} and for the muon $g-2$ anomaly in Section \ref{g-2}. {Our conclusions are outlined} in Section \ref{conclusions}.


\begin{figure}[]
\begin{center}
\includegraphics[scale=0.4]{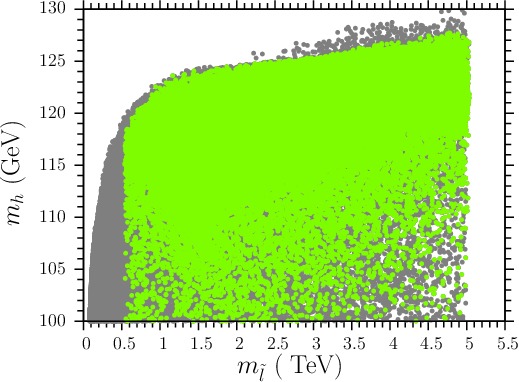}
\includegraphics[scale=0.4]{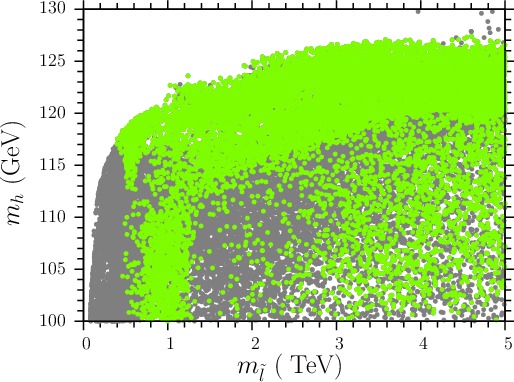}\vspace*{3mm}
\end{center}
\caption{Plots in the $m_h$ vs. $m_{\tilde{l}}$ plane for CMSSM (left panel) and NUHM2 (right panel).
{\it Gray} points are consistent with REWSB  and neutralino LSP.  {\it Green} points form a subset of the {\it gray} {points}
and satisfy the sparticle and Higgs mass bounds, {as well as} all other constraints described in Section \ref{pheno}.
}
\label{fig:1}
\end{figure}


\section{Phenomenological Constraints and Scanning Procedure \label{pheno}}

{We employ Isajet~7.84 \cite{ISAJET} interfaced with Micromegas 2.4 \cite{Belanger:2008sj} and FeynHiggs  2.10.0 \cite{feynhiggs}
to perform random scans over the parameter space.}
In Isajet, the weak scale values of gauge and third
generation Yukawa couplings are evolved to
$M_{\rm GUT}$ via the MSSM renormalization group equations (RGEs)
in the $\overline{DR}$ regularization scheme.
We do not strictly enforce the unification condition
$g_3=g_1=g_2$ at $M_{\rm GUT}$, since a few percent deviation
from unification can be assigned to unknown GUT-scale threshold
corrections~\cite{Hisano:1992jj}.
With the boundary conditions given at $M_{\rm GUT}$,
the SSB parameters, along with the gauge and third family Yukawa couplings,
are evolved back to the weak scale $M_{\rm Z}$.

In evaluating  the Yukawa couplings the SUSY threshold
corrections~\cite{Pierce:1996zz} are taken into account
at a common scale  $M_S= \sqrt{m_{\tilde t_L}m_{\tilde t_R}}$.
The entire parameter set is iteratively run between
$M_{\rm Z}$ and $M_{\rm GUT}$ using the full 2-loop RGEs
until a stable solution is obtained.
To better account for the leading-log corrections, one-loop step-beta
functions are adopted for the gauge and Yukawa couplings, and
the SSB scalar mass parameters $m_i$ are extracted from RGEs at appropriate scales
$m_i=m_i(m_i)$.The RGE-improved 1-loop effective potential is minimized
at an optimized scale  $M_S$, which effectively
accounts for the leading 2-loop corrections. Full 1-loop radiative
corrections are incorporated for all sparticle masses.

We implement the following random scanning procedure: A uniform and logarithmic distribution of random points is first generated in the given parameter space.
The function RNORMX \cite{Leva} is then employed
to generate a Gaussian distribution around each point in the parameter space.  The data points
collected all satisfy
the requirement of radiative electroweak symmetry breaking  (REWSB),
with the neutralino in each case being the LSP.

{ We use Micromegas to calculate the relic density and $BR(b \rightarrow s \gamma)$. The diphoton ratio $R_{\gamma \gamma}$ is calculated using FeynHiggs.} After collecting the data, we impose
the mass bounds on all the particles \cite{Nakamura:2010zzi} and use the
IsaTools package~\cite{Baer:2002fv}
to implement the various phenomenological constraints. We successively apply the following experimental constraints on the data that
we acquire from  ISAJET~7.84:
\begin{table}[h!]\centering
\begin{tabular}{rlc}
$123~{\rm GeV} \leq  m_h  \leq 127~{\rm GeV}$~~&\cite{:2012gk,:2012gu}&
\\
$ 0.8 \times 10^{-9} \ \leq \ BR(B_s \rightarrow \mu^+ \mu^-)  $&$ \leq\, 6.2 \times 10^{-9} \;
 (2\sigma)$        &   \cite{:2007kv}      \\
$2.99 \times 10^{-4} \ \leq  \ BR(b \rightarrow s \gamma) $&$ \ \leq\, 3.87 \times 10^{-4} \;
 (2\sigma)$ &   \cite{Barberio:2008fa}  \\
$0.15 \leq \frac{BR(B_u\rightarrow
\tau \nu_{\tau})_{\rm MSSM}}{BR(B_u\rightarrow \tau \nu_{\tau})_{\rm SM}}$&$ \leq\, 2.41 \;
(3\sigma)$. &   \cite{Barberio:2008fa}
\end{tabular}\label{table}
\end{table}

\section{Slepton Masses in CMSSM and NUHM2 \label{slepton}}

Before discussing the scenarios {where we} address the muon $g-2$ anomaly and  the decay rate $h\rightarrow \gamma \gamma$, we {first} present the {relationship} between {the} light CP even Higgs boson  and slepton masses in {two} well studied models, namely CMSSM and NUHM2. While it is true that radiative corrections to the  light CP even Higgs boson mass from {the} first two family sleptons are negligible, in the following section {we show} that relations among SSB mass terms {from} GUT scale boundary conditions in CMSSM and NUHM2 models {yield a} strong correlation {between them}.
 We do not consider the NUHM1 model since it is an intermediate step between CMSSM and NUHM2 in terms of the independent SSB parameters. Therefore, {the} light CP even Higgs boson mass dependence on slepton masses in NUHM1 {can be inferred, more or less, from the} CMSSM and NUHM2 models.

{We} have performed random scans in the fundamental parameter space of CMSSM and NUHM2 with ranges of the parameters given as follows:
\begin{flalign}
~~~~~~~~~~~~~~~~~~~~~~~~~~~~~~~~~~~&0 \leq  m_{16}  \leq 5\, \rm{TeV}& \nonumber  \\
~~~~~~~~~~~~~~~~~~~~~~~~~~~~~~~~~~~&0 \leq  M_{1/2}  \leq 3\, \rm{TeV} &\nonumber  \\
~~~~~~~~~~~~~~~~~~~~~~~~~~~~~~~~~~~&-3 \leq A_{0}/m_{3}  \leq 3 &\nonumber  \\
~~~~~~~~~~~~~~~~~~~~~~~~~~~~~~~~~~~&35 \leq  \tan\beta  \leq 55  ;& \nonumber
\end{flalign}
\vspace{-1cm}
\begin{flalign}
&{\rm For~ CMSSM:}~~~~~~~~~~~~~~~~~~ m_{16}=M_{H_u} = M_{H_d}& \nonumber \\
&{\rm For~~ NUMH2:}~~~~~~~~~ \ \ \  0 \leq  M_{H_u} \neq M_{H_d}  \leq 5\, \rm{TeV}&
\label{parameterRange}
\end{flalign}

\noindent
Here $m_{16}$ is the universal SSB mass parameter for sfermions, and $M_{1/2}$ denotes the universal SSB  gaugino masses. $A_{0}$ is the SSB trilinear scalar interaction coupling, $\tan\beta$ is the ratio of the MSSM Higgs vacuum expectation values (VEVs), and $M_{H_u}$, $M_{H_d}$ stand for  the SSB mass terms for  the MSSM up and down Higgs doublets. { Since the masses of the light CP even Higgs boson and sleptons do not change significantly  for $\tan\beta<35$, we used data from our former analysis for $35 \leq  \tan\beta  \leq 55$ to generate Figure \ref{fig:1}}.

In Figure \ref{fig:1} we display our results in the $m_h - m_{\tilde{l}}$ plane for CMSSM (left panel) and NUHM2 (right panel). Here $m_{\tilde{l}}$ stands for {the} left handed slepton masses {for the first two families}. We {observe} that in the CMSSM and NUHM2 models there is a {fairly} strong correlation between the Higgs boson mass ($m_h$) and  the first two generation slepton masses ($m_{\tilde{l}}$).  Note that the bounds for {the} right handed slepton masses are very similar {and are therefore not displayed}.
{The} {\it gray} points are consistent with REWSB  and neutralino LSP, {and the} {\it green} points form a subset of the {\it gray} {points}
and satisfy the sparticle and Higgs mass bounds, {as well as} all other constraints described in Section \ref{pheno}.

We see from Figure \ref{fig:1} that for both the CMSSM and NUHM2 models, {compatibility with the measurement} $123~{\rm GeV} \leq m_h \leq 127~{\rm GeV}$ requires that the slepton masses lie above 1 TeV. The salient features of the results in Figure 1 can be understood by noting that in order for the stop quark
mass to be more than 1 TeV \cite{Ajaib:2012vc} (which is necessary to achieve $m_h \approx 125$ GeV), with universal SSB parameters $M_{1/2}$ and $m_0$, the first  and second generation
squark masses {acquire masses} in the  multi-TeV range, and the corresponding smuon masses lie
around the TeV scale. On the other hand, as mentioned above, in order to have an enhancement in muon $g-2$ and in the decay rate of  $h\rightarrow \gamma \gamma$, the sleptons need to be much lighter than 1 TeV.
Overall, we learn from Figure 1 that in the CMSSM, NUHM1 and NUHM2 scenarios, it is not possible to have enhancement in muon $g-2$ and the decay rate of $h\rightarrow \gamma \gamma$ {relative to the Standard Model}. This conclusion motivates us to explore {alternative} scenarios
which can simultaneously accommodate an enhancement or suppresion of  $h\rightarrow \gamma \gamma$ and an enhancement in muon $g-2$.

\begin{figure}[]
\centering
{{\includegraphics[scale=0.4]{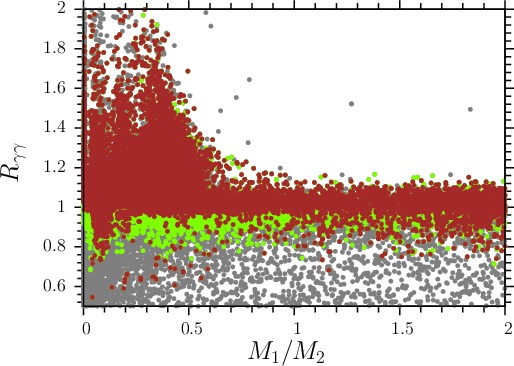}}}\hfill
{{\includegraphics[scale=0.4]{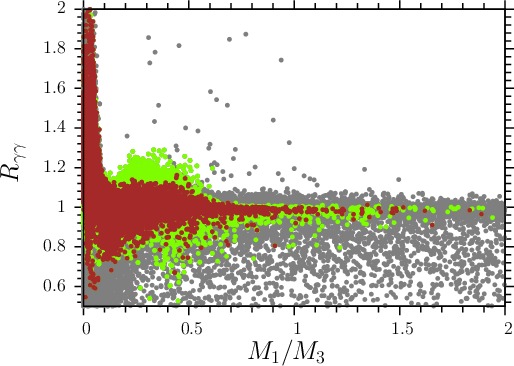}}}\\
{{\includegraphics[scale=0.4]{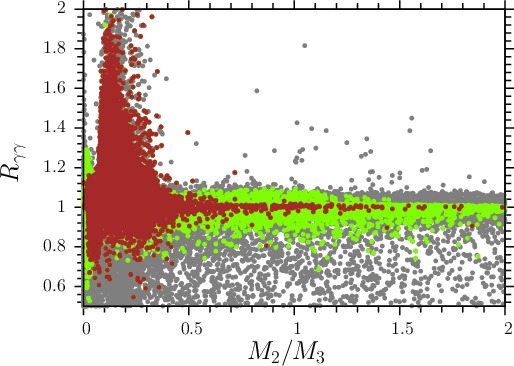}}}\hfill
{{\includegraphics[scale=0.4]{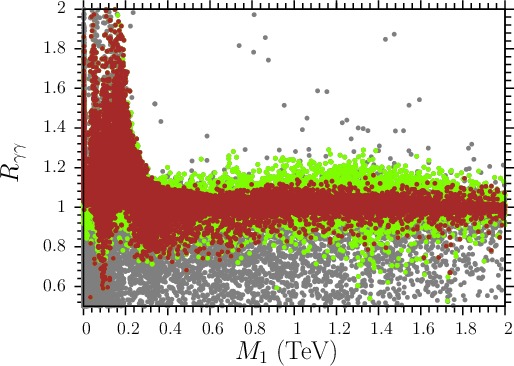}}}\\
{{\includegraphics[scale=0.4]{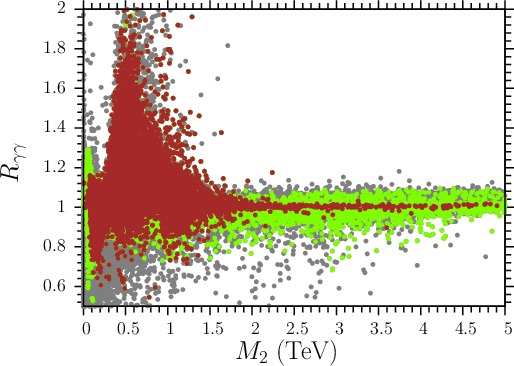}}}\hfill
{{\includegraphics[scale=0.4]{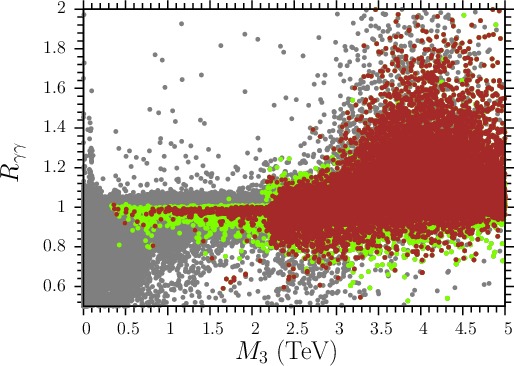}}}
\caption{Plots in the $R_{\gamma \gamma} - M_1/M_2$, $R_{\gamma \gamma} - M_1/M_3$, $R_{\gamma \gamma} - M_2/M_3$, $R_{\gamma \gamma} - M_1$, $R_{\gamma \gamma} - M_2$ and $R_{\gamma \gamma} - M_3$ {planes}. {\it Gray} points are consistent with REWSB and neutralino LSP. {\it Green} points form a
subset of the {\it gray} points and satisfy the sparticle and Higgs mass bounds, as well
as all other constraints described in Section \ref{pheno}.
{\it Brown}  points belong to a
subset of {\it green} points and satisfy {the following} bound on the LSP neutralino relic abundance,
$0.001 \leq \Omega h^2 \leq 1$.}
\label{fig:2}
\end{figure}


\section{$h\rightarrow \gamma \gamma$ Decay and Particle Spectra\label{Rgg}}


One of the most promising Higgs boson decay channels is the $\gamma \gamma$ final state which, at leading order, proceeds through a loop containing charged particles, {including} the charged Higgs, sfermions and charginos.{ In the SM, the leading contribution to $h \rightarrow \gamma \gamma$ decay comes from the W boson loop, the top loop being the next dominant one.} The decay width is given by (see \cite{Djouadi:2005gj,gunion90} and references therein)
\begin{eqnarray}
\Gamma(h\rightarrow\gamma\gamma)&=&\frac{G_{F}\alpha^{2}m_{h}^{3}}{128\sqrt{2}\pi}
\left| N_{c}\, Q_{t}^{2}\, g_{htt}\, A_{1/2}^{h}(\tau_{t})+
g_{hWW}\, A_{1}^{h}(\tau_{W}) + {\cal A}_{  \text{\tiny{SUSY}}  }^{\gamma\gamma}\right|^2,
\end{eqnarray}
where $g_{hWW}$ is the coupling of $h$ to the $W$ boson. The supersymmetric contribution $ {\cal A}_{  \text{\tiny{SUSY}}  }^{\gamma\gamma} $ is given by
\begin{eqnarray}
{\cal A}_{  \text{\tiny{SUSY}}  }^{\gamma\gamma} &=&
 g_{hH^{+}H^{-}}\, \frac{m_{W}^{2}}{m^{2}_{H^{\pm}}} \, A_{0}^{h}(\tau_{H^{\pm}}) +
\sum_f N_c Q_f^2\, g_{h\tilde{f}\tilde{f}}\, \frac{m_Z^2}{m^2_{\tilde{f}}}\, A_0^h(\tau_{\tilde{f}})
     + \nonumber \\
&& \quad \quad \quad \sum_i g_{h\chi_i^+\chi_i^-}\, \frac{m_W}{m_{\chi_i}}\, A_{\frac12}^h(\tau_{\chi_i}),
\end{eqnarray}
where $g_{hXX}$ is the coupling of $h$ to the particle $X$ ($= H^{\pm}, \tilde{f}, \chi^{\pm}_i$).

The stop and sbottom loop factors have similar contributions as the gluon fusion case. In this case, however, the stau can also contribute to enhance the decay width without changing the gluon fusion cross section. The chargino contribution to the decay width is known to be less than 10\% for $m_{\chi_{i}^{\pm}} \gtrsim 100 \rm \ GeV$. The charged Higgs contribution is even smaller since its coupling to the CP-even Higgs is not proportional to its mass and also due to the loop suppression $m_{W}^{2}/m^{2}_{H^{\pm}}$.

In the MSSM framework it was shown \cite{Carena:2012xa}  that only a light stau can give significant enhancement/suppresion  in the {process} $gg \rightarrow h  \rightarrow \gamma \gamma$,
while keeping the lightest CP-even Higgs boson mass in the interval $123~{\rm GeV} \leq m_h \leq 127~{\rm GeV}$.

In this paper  we discuss the  {scenario  with non-universal and opposite sign  gaugino {masses at $M_{\rm GUT}$,   {while} the sfermion masses at $M_{GUT}$ assumed to be universal}. This is a follow up of the work presented in ref. \cite{Gogoladze:2014cha},  where it was shown that
 the muon $g-2$ anomaly can be explained in this model, but the decay rate for $h \rightarrow \gamma \gamma$ was not analyzed. It was shown in ref. \cite{Gogoladze:2014cha} that the sleptons can be as light as 100 GeV in this model.
 This observation motivated us to investigate the decay rate for $h \rightarrow \gamma \gamma$ and study the parameter space {which yields} enhancement or suppression for this {process}.

We  { perform random scans for} following ranges of the parameters:
\begin{eqnarray}
0 \leq  m_{16}  \leq 3\, \rm{TeV} \nonumber  \\
0 \leq  M_{1}  \leq 5\, \rm{TeV} \nonumber  \\
0 \leq  M_{2}  \leq 5\, \rm{TeV} \nonumber  \\
-5 \leq  M_{3}  \leq 0\, \rm{TeV} \nonumber  \\
-3 \leq A_{0}/m_{16}  \leq 3 \nonumber  \\
2 \leq  \tan\beta  \leq 60 \nonumber \\
0 \leq  m_{10}  \leq 5\, \rm{TeV} \nonumber \\
\mu > 0.
\label{parameterRange}
\end{eqnarray}
Here  $M_{1}$, $M_{2}$, and $M_{3}$ denote the SSB gaugino masses for $U(1)_{Y}$, $SU(2)_{L}$ and $SU(3)_{c}$ respectively. We choose different sign for gauginos which was again motivated from the work presented in ref \cite{Gogoladze:2014cha}, where it was shown that an opposite sign non-universal gaugino mass case is more preferable from the muon $g-2$ point of view than the same sign non-universal gaugino case.

The main message of Section \ref{slepton} is that {with} universal SSB mass terms for the gaugino and sfermion sectors, it is impossible to have significant SUSY contributions to the decay $h \rightarrow \gamma \gamma $ and muon $g-2$.  On the other hand, as shown in ref. \cite{Gogoladze:2014cha}, non-universal gaugino masses allow for {sufficiently} light sleptons while keeping the colored sparticles in the multi TeV region. Because of this observation, we investigate the extent to which non-universality is allowed in the gaugino sector to enhance or suppress the decay channel $h \rightarrow \gamma \gamma $.   The color coding in Figure \ref{fig:2} is given as follows, {\it Gray} points are consistent with REWSB and neutralino LSP. {\it Green} points form a
subset of the {\it gray} points and satisfy the sparticle and Higgs mass bounds, as well
as all other constraints described in Section \ref{pheno}.
{\it Brown}  points belong to a
subset of {\it green} points and satisfy {the constraint} $0.001 \leq \Omega h^2 \leq 1$ on the LSP neutralino relic abundance. We have chosen to display our results for a wider range
of $\Omega h^2$, keeping in mind that one can always find points compatible with
the current WMAP range for relic abundance with dedicated scans within the brown
regions.

\begin{figure}[]
\centering
{\label{fig:3}{\includegraphics[scale=0.4]{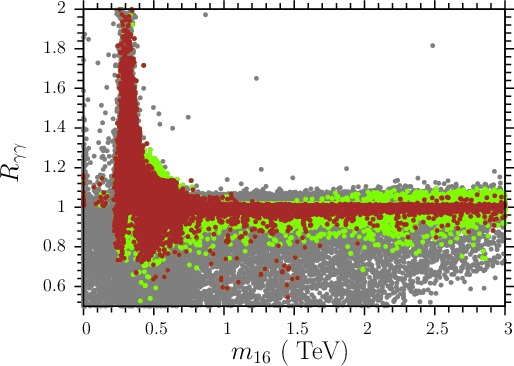}}}\hfill
{\label{fig:3}{\includegraphics[scale=0.4]{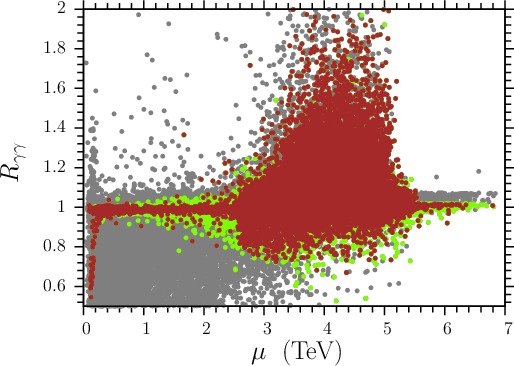}}}\\
{\label{fig:3}{\includegraphics[scale=0.4]{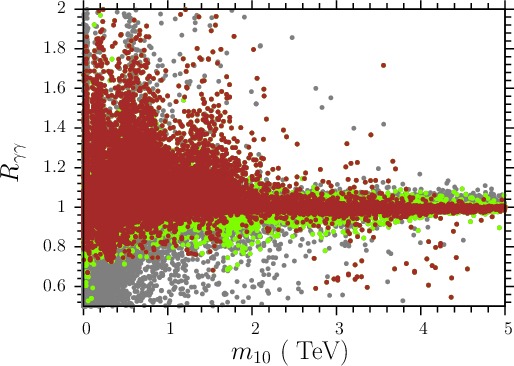}}}\hfill
{\label{fig:3}{\includegraphics[scale=0.4]{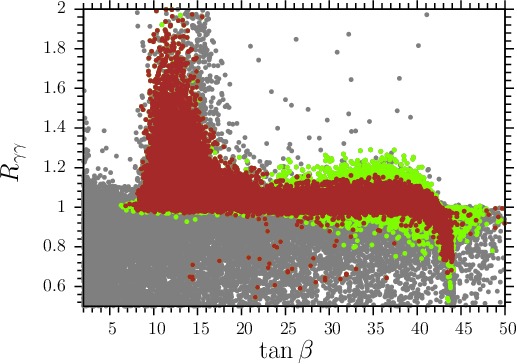}}}
\caption{Plots in the $R_{\gamma \gamma} - m_{16}$, $R_{\gamma \gamma} - \mu$, $R_{\gamma \gamma} - m_{10}$ and $R_{\gamma \gamma} - \tan\beta$ planes. Color coding  same as in  Figure 2.}
\label{fig-3}
\end{figure}


The results from the  $R_{\gamma \gamma} - M_1/M_2$, $R_{\gamma \gamma} - M_1/M_3$ and $R_{\gamma \gamma} - M_2/M_3$ planes show that a significant deviation from universality of gaugino masses in order to have sizable SUSY contribution to $h \rightarrow \gamma \gamma$ decay {is necessary}. For instance, the ratio $M_1/M_3$ needs to be more than 5, while $M_2/M_3> 3$ and  $M_1/M_2 > 2$. Not only do we observe a strict prediction of gaugino mass ratios, but also a precise prediction of their values. {In particular, from the } $R_{\gamma \gamma} - M_1$ panel {we can see} that it is difficult to have {an} enhancement of $h\rightarrow \gamma \gamma$  if $M_1 \gtrsim 300$ GeV. At the same time, the upper bound on $M_2$ is {less stringent} and enhancement {of} $h\rightarrow \gamma \gamma$ occurs even with $M_2$  around 1 TeV.

{ We observe from the $R_{\gamma \gamma} - M_3$ panel that the  parameter $M_{3} \gtrsim 2.5$ TeV. The reason for such a large value of $M_3$ is the following.  Since we assume universality in sfermion masses and seek solutions {with} sleptons {not heavier} than a few hundred GeV, $m_{16}$ is required to be around a few hundred GeV. Moreover, {with a} tau slepton {mass of} around a hundred GeV, in order to avoid breaking {the} charge symmetry, $A_{\tau}$ needs to be around a hundred GeV. This places a constraint on $A_t$ because we assume a universal trilinear SSB $A_0$ term.
Consequently, a relatively small value of $A_t$ is obtained at low scale. On the other hand, it was shown in \cite{Ajaib:2012vc} that the stop mass needs to be more than 3 TeV {if} $A_t$ is not the dominant contributor to the radiative corrections of the light CP-even Higgs boson mass. In order to obtain such a heavy stop quark, when $m_{16}$ is of order hundred GeV, a {fairly} large $M_3$ is required. This tendency can be observed from the {following} semi-analytic {expressions} for stop quark {masses} \cite{Gogoladze:2009bd}
\begin{eqnarray}
m_{Q_{t}}^{2}
& \approx &5.41 M_{3}^{2}+0.392 M_{2}^{2}+0.64 m_{16}^{2}
+0.115 M_{3}A_{t_{0}}
-0.072 M_{3}M_{2}+ \ldots, \nonumber
\\
m_{U_{t}}^{2}
& \approx &4.52 M_{3}^2-0.188 M_{2}^2+0.273 m_{16}^2-0.066 A_{t_0}^2  -0.145 M_{3}M_{2} + \ldots.
\label{semi}
 \end{eqnarray}
It is clear from Eq. (\ref{semi}) that {if} $m_{16}$, $M_2$ and $A_t$ are of the order {of} hundred GeV or so, the way to obtain several TeV stop quark masses is to also have $M_3$ around several TeV.
}

In Figure \ref{fig-3} we display our results for the fundamental parameters in the $R_{\gamma \gamma} - m_{16}$, $R_{\gamma \gamma} - \mu$, $R_{\gamma \gamma} - m_{10}$ and $R_{\gamma \gamma} - \tan\beta$ planes. We can see that an enhancement in the $h\rightarrow \gamma \gamma$ channel constrains the parameters in this model. The fundamental parameter $m_{16}$ is restricted to a narrow range, $200 {\rm \ GeV} \lesssim m_{16} \lesssim 600  {\rm \ GeV}$, for  $R_{\gamma \gamma}\ge 1.1$. Similarly, the range for the other parameters for a corresponding enhancement in the $h\rightarrow \gamma \gamma$ channel are:
$2.5 {\rm \ TeV} \lesssim \mu \lesssim 5.5  {\rm \ TeV}$,
$m_{10} \lesssim 2  {\rm \ TeV}$,
$10 \lesssim \tan\beta \lesssim 20$.

Figure \ref{fig:4} shows our results for the sparticle masses in the $R_{\gamma \gamma} - m_{\tilde{\tau}}$, $R_{\gamma \gamma} - m_{\tilde{\chi}^0_1}$, $R_{\gamma \gamma} - m_{\tilde{g}}$ and $R_{\gamma \gamma} - m_{\tilde{t}_1}$ planes. For $R_{\gamma \gamma}\ge 1.1$, the stau and the neutralino  are both {relatively} light with mass ranges $100 {\rm \ GeV} \lesssim m_{\tilde{\tau}} \lesssim 200 {\rm \ GeV} $ and $50 {\rm \ GeV} \lesssim m_{\tilde{\chi}_1^0} \lesssim 200 {\rm \ GeV}$. From the lower panels of Figure \ref{fig:4} we can see that the colored sparticles corresponding to $R_{\gamma \gamma}\ge 1.1$ are heavy with $ m_{\tilde{g}} \gtrsim 4.5 {\rm \ TeV} $ and $ m_{\tilde{t}_1} \gtrsim 3.5 {\rm \ TeV} $.
{ The reason for such heavy stop and gluino masses has been discussed above.}
Testing squarks and gluinos with this mass would be challenging at 14 TeV LHC.

\begin{figure}[]

\centering
{\label{fig:4}{\includegraphics[scale=0.4]{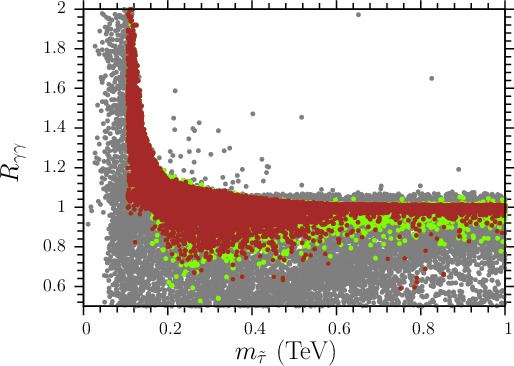}}}\hfill
{\label{fig:4}{\includegraphics[scale=0.4]{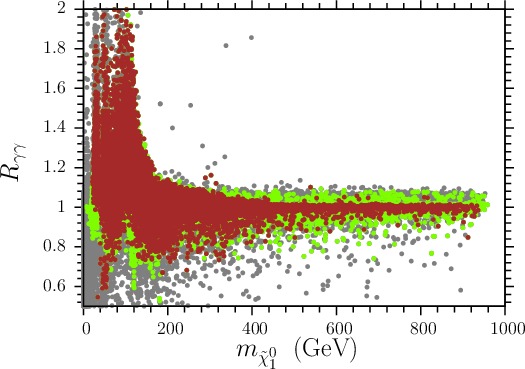}}}\\
{\label{fig:4}{\includegraphics[scale=0.4]{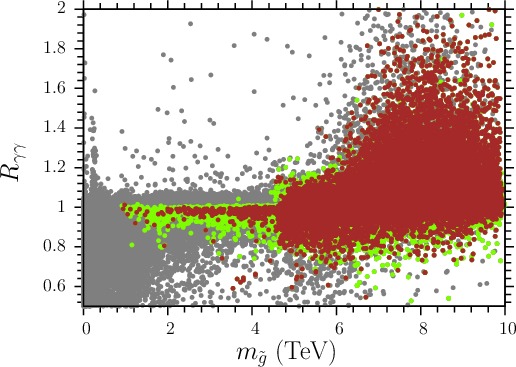}}}\hfill
{\label{fig:4}{\includegraphics[scale=0.4]{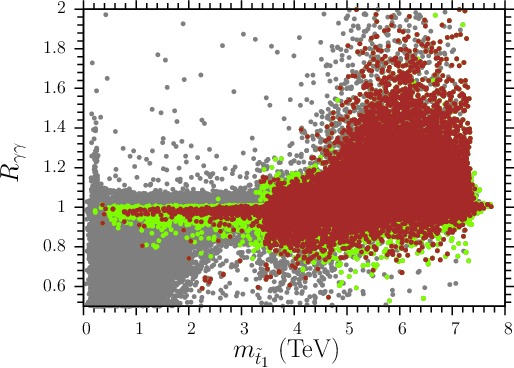}}}\\
\caption{Plots in the $R_{\gamma \gamma} - m_{\tilde{\tau}}$, $R_{\gamma \gamma} - m_{\tilde{\chi}^0_1}$, $R_{\gamma \gamma} - m_{\tilde{g}}$ and $R_{\gamma \gamma} - m_{\tilde{t}_1}$ planes. Color coding  same as in Figure 2.}
\label{fig:4}
\end{figure}


\begin{figure}[]

\centering
{\label{fig:5}{\includegraphics[scale=0.4]{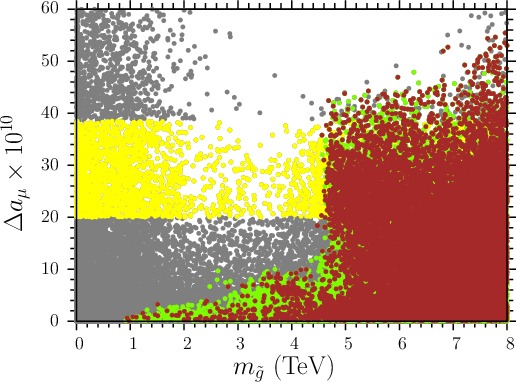}}}\hfill
{\label{fig:5}{\includegraphics[scale=0.4]{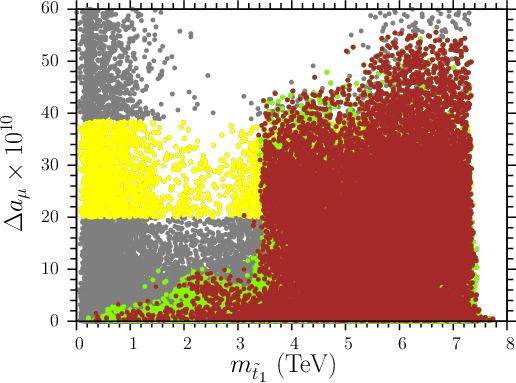}}}
\caption{Plots in the $\Delta a_\mu  - m_{\tilde{g}}$ and $\Delta a_\mu \times 10^{10} - m_{\tilde{t}_1}$ planes. Color coding  same as in Figure 2.}
\label{fig:5}
\end{figure}


\begin{figure}[]

\centering
{\label{fig:6}{\includegraphics[scale=0.4]{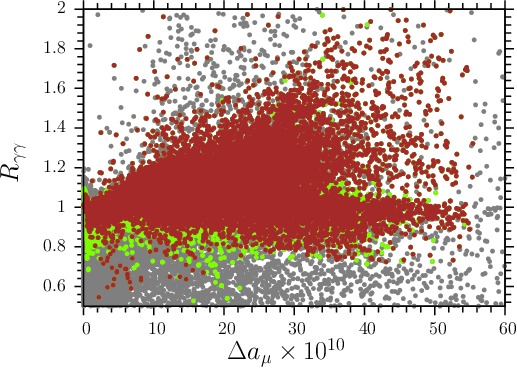}}}
\caption{Plot in the the $R_{\gamma \gamma} -\Delta a_\mu $ plane. Color coding  same as in Figure 2.}
\label{fig:6}
\end{figure}


\section{\label{g-2}The Muon Anomalous Magnetic Moment}

The leading  contribution from low scale supersymmetry  to the muon anomalous magnetic moment
\cite{Moroi:1995yh} depends on the following parameters:
\begin{equation}
 M_1, \, M_2, \, \mu,\, \tan\beta,  m_{\tilde{\mu}_{L}}, \, m_{\tilde{\mu}_{R}},
\label{eqq3}
\end{equation}
Since we assume a universal the trilinear SSB term $A_0$, it follows that $A_\mu < \mu \tan\beta$ and we therefore do not consider the trilinear SSB-term contribution here.

 The colored particles do not directly provide significant contribution to the muon $g-2$ calculation but are still constrained from the bound on the light CP even Higgs boson mass and the  muon $g-2$ calculation. Figure \ref{fig:5} shows our results in the $\Delta a_\mu \times 10^{10} - m_{\tilde{g}}$ and $\Delta a_\mu \times 10^{10} - m_{\tilde{t}_1}$ planes.
We can see that the above mentioned constraint yields a {stringent} lower bound for the  gluino and stop {masses}. Both these sparticles have to be heavier than 3 TeV, which makes it very hard  {to see them} in the second LHC run. However, there is hope that these sparticles can be observed at {a future} 100 TeV collider.

\begin{table}
\centering
\begin{tabular}{|p{2.6cm}|p{2.6cm}p{2.6cm}p{2.6cm}p{2.6cm}|}
\hline
\hline
                 	&	 Point 1 	&	 Point 2 	&	 Point 3 	&	 Point 4 	\\
\hline									
									
									
$m_{16}$         	&$	351	$&$	438	$&$	561	$&$	392	$\\
$m_{10}$         	&$	451	$&$	32	$&$	478	$&$	2.7	$\\
$A_0/m_{16}$         	&$	-2.6	$&$	1.6	$&$	0.3	$&$	2.6	$\\
$\tan\beta$      	&$	12	$&$	26	$&$	42	$&$	43	$\\
$M_{1}$         	&$	88	$&$	5	$&$	344	$&$	749	$\\
$M_{2} $         	&$	714	$&$	1051	$&$	124	$&$	722	$\\
$M_{3} $         	&$	-4913	$&$	-4550	$&$	-4420	$&$	-4524	$\\
		  		  		  		  	
\hline		  		  		  		  	
$\mu$            	&$	471	$&$	747	$&$	680	$&$	811	$\\

\hline		  		  		  		  	
$m_h$            	&$	123	$&$	124	$&$	124	$&$	125	$\\
$m_H$            	&$	4847	$&$	4074	$&$	1607	$&$	591	$\\
$m_A$            	&$	4815	$&$	4047	$&$	1596	$&$	587	$\\
$m_{H^{\pm}}$    	&$	4847	$&$	4075	$&$	1610	$&$	600	$\\
		  		  		  		  	
\hline		  		  		  		  	
$m_{\tilde{\chi}^0_{1,2}}$	&$	          75,          720	$&$	          33,          999	$&$	         184,          198	$&$	         367,          712	$\\

$m_{\tilde{\chi}^0_{3,4}}$	&$	        4774,         4774	$&$	        4570,         4570	$&$	        4512,         4513	$&$	        4628,         4628	$\\

$m_{\tilde{\chi}^{\pm}_{1,2}}$	&$	         726,         4729	$&$	        1004,         4528	$&$	         199,         4470	$&$	         714,         4585	$\\

$m_{\tilde{g}}$  	&$	9709	$&$	9029	$&$	8857	$&$	9008	$\\
		  		  		  		  	
\hline $m_{ \tilde{u}_{L,R}}$	&$	        8247,         8262	$&$	        7703,         7697	$&$	        7553,         7582	$&$	        7674,         7689	$\\
                 		  		  		  		  	
$m_{\tilde{t}_{1,2}}$	&$	        7217,         7800	$&$	        6664,         7146	$&$	        6573,         6707	$&$	        6625,         6754	$\\
                 		  		  		  		  	
\hline $m_{ \tilde{d}_{L,R}}$	&$	        8247,         8269	$&$	        7704,         7703	$&$	        7553,         7588	$&$	        7675,         7692	$\\
                 		  		  		  		  	
$m_{\tilde{b}_{1,2}}$	&$	        7756,         8208	$&$	        7107,         7432	$&$	        6613,         6750	$&$	        6608,         6750	$\\
                 		  		  		  		  	
\hline		  		  		  		  	
$m_{\tilde{\nu}_{1}}$	&$	410	$&$	727	$&$	446	$&$	543	$\\
                 		  		  		  		  	
$m_{\tilde{\nu}_{3}}$	&$	419	$&$	728	$&$	677	$&$	813	$\\
                 		  		  		  		  	
\hline		  		  		  		  	
$m_{ \tilde{e}_{L,R}}$	&$	         561,          150	$&$	         787,          354	$&$	         469,          549	$&$	         552,          449	$\\
                		  		  		  		  	
$m_{\tilde{\tau}_{1,2}}$	&$	         176,          519	$&$	         160,          786	$&$	         368,          892	$&$	         526,          979	$\\
                		  		  		  		  	
\hline		  		  		  		  	
$\Delta(g-2)_{\mu}$  	&$	  29.5\times 10^{-10}	$&$	  4.67\times 10^{-10}	$&$	  29.3\times 10^{-10}	$&$	  29.4\times 10^{-10}	$\\

$R_{\gamma \gamma}$	&$	1.2	$&$	1.1	$&$	1	$&$	0.74	$\\

\hline
\end{tabular}
\caption{{Four} benchmark points from our analysis. The first point has $R_{\gamma \gamma}=1.2$ with the central value of $g-2$. The second point has $(g-2)_\mu$ {consistent with} the SM {and} $R_{\gamma \gamma}=1.1$. The third point also has the central value of $(g-2)_{\mu}$ but with no enhancement in the diphoton channel. {The fourth point shows a suppression in the diphoton channel with $R_{\gamma \gamma}=0.74$}. For all these points the sleptons are {relatively} light {but} the colored sparticles are considerably {heavier}.}
\label{tab1}
\end{table}

In Figure \ref{fig:6} we display {results in the} $R_{\gamma \gamma} -\Delta a_\mu $ plane. The plot shows that there is a considerable region of the parameter space that allows for simultaneous enhancement in the decay channel $h \rightarrow \gamma\gamma$ and muon $g-2$. In this model we can have the correct neutralino dark matter relic abundance through the slepton coannihilation channel. It is also interesting that this model connects the parameter space relevant for two different experiments. If the enhancement in the $h \rightarrow \gamma\gamma$ decay channel is excluded than a considerable region of the parameter space that explains the $g-2$ anomaly will also be excluded.

Finally, in Table \ref{tab1} we display {four} benchmark points from our analysis. All these points satisfy the constraints described in section 2. The first and third point {yield a muon $g-2$} around the central {measured} value with  $R_{\gamma \gamma} = 1.2$ and  $R_{\gamma \gamma} = 1.0$, respectively. The second point has $(g - 2 )_\mu$ {consistent with} the SM and  $R_{\gamma \gamma} = 1.1$. {The fourth point shows a suppression in the diphoton channel with $R_{\gamma \gamma}=0.74$}. For all these points the sleptons are light {while} the colored sparticles are considerably heavier.

\section{ Conclusion \label{conclusions}}

We studied a supersymmetric model with non-universal gaugino masses at $M_{GUT}$ that accommodates enhancement or suppression in the $h\rightarrow \gamma \gamma$ channel while simultaneously explaining the muon $g-2$ anomaly. The parameter space {we obtain} is consistent with the current bounds on the sparticle masses and constraints from B-physics. {The desired} neutralino dark matter relic abundance is achieved in this model through slepton coannihilation channel. We find that the parameter space with $R_{\gamma \gamma}\ge 1.1$  predicts {relatively} light sleptons with a stau mass range of  $100 {\rm \ GeV} \lesssim m_{\tilde{\tau}} \lesssim 200 {\rm \ GeV} $. The colored sparticles corresponding to $R_{\gamma \gamma}\ge 1.1$ are heavy with $ m_{\tilde{g}} \gtrsim 4.5 {\rm \ TeV} $ and $ m_{\tilde{t}_1} \gtrsim 3.5 {\rm \ TeV} $, which makes it very challenging to {observe them in} the second run of LHC.


\section*{Acknowledgments}

This work is supported in part by the DOE Grant No. DE-FG02-12ER41808.  This work used the Extreme Science
and Engineering Discovery Environment (XSEDE), which is supported by the National Science
Foundation grant number OCI-1053575. I.G. acknowledges support from the  Rustaveli
National Science Foundation  No. 03/79.



\begin{thebibliography}{99}
\bibitem{:2012gk}
  G.~Aad {\it et al.}  [ATLAS Collaboration],
  Phys.\ Lett.\ B {\bf 716}, 1 (2012).

\bibitem{:2012gu}
  S.~Chatrchyan {\it et al.}  [CMS Collaboration],
  Phys.\ Lett.\ B {\bf 716}, 30 (2012).

\bibitem{Martin:1997ns}
 See, for instance, S.~P.~Martin,
  arXiv:hep-ph/9709356 [hep-ph] and references therein.


\bibitem{Aad:2012fqa}
  G.~Aad {\it et al.}  [ATLAS Collaboration],
  Phys.\ Rev.\ D {\bf 87}, 012008 (2013).


\bibitem{Chatrchyan:2012jx}
  S.~Chatrchyan {\it et al.}  [CMS Collaboration],
  JHEP {\bf 1210}, 018 (2012).

\bibitem{Gogoladze:2009bn}
  I.~Gogoladze, R.~Khalid and Q.~Shafi,
  Phys.\ Rev.\ D {\bf 80}, 095016 (2009);
%
  M.~A.~Ajaib, T.~Li and Q.~Shafi,
  Phys.\ Lett.\ B {\bf 705}, 87 (2011);
  S.~Raza, Q.~Shafi and C.~S.~Ün,
  arXiv:1412.7672 [hep-ph].





\bibitem{Kane:1993td}
  G.~L.~Kane, C.~F.~Kolda, L.~Roszkowski and J.~D.~Wells,
  Phys.\ Rev.\ D {\bf 49}, 6173 (1994).

\bibitem{Baer:2004fu}
  H.~Baer, A.~Mustafayev, S.~Profumo, A.~Belyaev and X.~Tata,
  Phys.\ Rev.\ D {\bf 71}, 095008 (2005).





  \bibitem{nuhm2} J. Ellis, K. Olive and Y. Santoso, \plb{539}{2002}{107};
  J.~Ellis, T.~Falk, K.~Olive and Y.~Santoso, \npb{652}{2003}{259};
  H.~Baer, A.~Mustafayev, S.~Profumo, A.~Belyaev and X.~Tata, \jhep{0507}{2005}{065}.



  \bibitem{Davier:2010nc}
  M.~Davier, A.~Hoecker, B.~Malaescu and Z.~Zhang,
  Eur.\ Phys.\ J.\ C {\bf 71}, 1515 (2011)  [Erratum-ibid.\ C {\bf 72}, 1874 (2012)];
   K.~Hagiwara, R.~Liao, A.~D.~Martin, D.~Nomura and T.~Teubner,
  J.\ Phys.\ G {\bf 38}, 085003 (2011).




\bibitem{Bennett:2006fi}
  G.~W.~Bennett {\it et al.}  [Muon (g-2) Collaboration],
  Phys.\ Rev.\ D {\bf 73}, 072003 (2006);
  Phys.\ Rev.\ D {\bf 80}, 052008 (2009).

\bibitem{Moroi:1995yh}
  T.~Moroi,
  Phys.\ Rev.\ D {\bf 53}, 6565 (1996)
  [Erratum-ibid.\ D {\bf 56}, 4424 (1997)];
  S.~P.~Martin and J.~D.~Wells,
  Phys.\ Rev.\ D {\bf 64}, 035003 (2001).
  G.~F.~Giudice, P.~Paradisi, A.~Strumia and A.~Strumia,
  JHEP {\bf 1210}, 186 (2012)




\bibitem{Gogoladze:2014cha}
  I.~Gogoladze, F.~Nasir, Q.~Shafi and C.~S.~Un,
  Phys.\ Rev.\ D {\bf 90}, 035008 (2014).



\bibitem{Akula:2013ioa}
  S.~Mohanty, S.~Rao and D.~P.~Roy,
  JHEP {\bf 1309}, 027 (2013);
  S.~Akula and P.~Nath,
  Phys.\ Rev.\ D {\bf 87}, 115022 (2013);
  J.~Chakrabortty, S.~Mohanty and S.~Rao,
  arXiv:1310.3620 [hep-ph].




\bibitem{Babu:2014sga}
  K.~S.~Babu, I.~Gogoladze, S.~Raza and Q.~Shafi,
  Phys.\ Rev.\ D {\bf 90}, 056001 (2014).

\bibitem{Ibe:2013oha}
  M.~Ibe, T.~T.~Yanagida and N.~Yokozaki,
  JHEP {\bf 1308}, 067 (2013).










  \bibitem{Ajaib:2014ana}
  M.~A.~Ajaib, I.~Gogoladze, Q.~Shafi and C.~S.~Un,
  JHEP {\bf 1405}, 079 (2014).





\bibitem{yukawaUn}
B. Ananthanarayan, G. Lazarides and Q. Shafi, Phys. Rev. D {\bf 44},
1613 (1991);  Phys.\ Lett.\ B {\bf 300}, 245 (1993); Q.~Shafi and
B.~Ananthanarayan, Proceedings of the Summer School in High Energy Physics and Cosmology 1991, edited by E. Gava, K. Narain,  S.  Randjbar-Daemi,  E.  Sezgin,  and  Q.  Shafi, ICTP Series in Theoretical Physics Vol. 8 (World Scientific, River Edge, NJ, 1992).

\bibitem{Baer:2004xx}
  H.~Baer, A.~Belyaev, T.~Krupovnickas and A.~Mustafayev,
  JHEP {\bf 0406}, 044 (2004);
  K.~S.~Babu, I.~Gogoladze, Q.~Shafi and C.~S.~Un,
  Phys.\ Rev.\ D {\bf 90}, 116002 (2014).




\bibitem{Carena:2012xa}
  M.~Carena, I.~Low and C.~E.~M.~Wagner,
  JHEP {\bf 1208}, 060 (2012)
  K.~Schmidt-Hoberg, F.~Staub and M.~W.~Winkler,
  JHEP {\bf 1301} (2013) 124;
  M.~Carena, S.~Gori, N.~R.~Shah and C.~E.~M.~Wagner,
  JHEP {\bf 1203} (2012) 014;
  M.~A.~Ajaib, I.~Gogoladze and Q.~Shafi,
  Phys.\ Rev.\ D {\bf 86}, 095028 (2012)
  N.~Maru and N.~Okada,
  Phys.\ Rev.\ D {\bf 87}, no. 9, 095019 (2013);
  J.~Guo, Z.~Kang, J.~Li and T.~Li,
  arXiv:1308.3075 [hep-ph].
  M.~Hemeda, S.~Khalil and S.~Moretti,
  Phys.\ Rev.\ D {\bf 89}, no. 1, 011701 (2014).
  A.~Chakraborty, B.~Das, J.~L.~Diaz-Cruz, D.~K.~Ghosh, S.~Moretti and P.~Poulose,
  Phys.\ Rev.\ D {\bf 90}, no. 5, 055005 (2014);
  S.~Chakraborty, A.~Datta and S.~Roy,
  arXiv:1411.1525 [hep-ph].







  \bibitem{ATLAS-mugg-1}
  G.~Aad {\it et al.}  [ ATLAS Collaboration],
``Measurement of Higgs boson production in the diphoton decay channel
  in $pp$ collisions at center-of-mass energies of 7 and 8 TeV with
  the ATLAS detector,''
  arXiv:1408.7084 [hep-ex].


\bibitem{CMS-mugg-2}
  V.~Khachatryan {\it et al.}  [CMS Collaboration],
  ``Observation of the diphoton decay of the Higgs boson and
  measurement of its properties,''
  arXiv:1407.0558 [hep-ex].









\bibitem{ISAJET}
  F.~E.~Paige, S.~D.~Protopopescu, H.~Baer and X.~Tata,
  hep-ph/0312045.

\bibitem{Belanger:2008sj}
  G.~Belanger, F.~Boudjema, A.~Pukhov and A.~Semenov,
  Comput.\ Phys.\ Commun.\  {\bf 180}, 747 (2009).



\bibitem{feynhiggs}
  M.~Frank, T.~Hahn, S.~Heinemeyer, W.~Hollik, H.~Rzehak and G.~Weiglein,
  JHEP {\bf 0702}, 047 (2007);
%
  G.~Degrassi, S.~Heinemeyer, W.~Hollik, P.~Slavich and G.~Weiglein,
  Eur.\ Phys.\ J.\ C {\bf 28}, 133 (2003);
%
  S.~Heinemeyer, W.~Hollik and G.~Weiglein,
  Eur.\ Phys.\ J.\ C {\bf 9}, 343 (1999);
%
  S.~Heinemeyer, W.~Hollik and G.~Weiglein,
  Comput.\ Phys.\ Commun.\  {\bf 124}, 76 (2000).



\bibitem{Hisano:1992jj}
J.~Hisano, H.~Murayama  , and T.~Yanagida,
  { Nucl. Phys.} {\bf B402} (1993) 46.
Y.~Yamada,
{ Z. Phys.} {\bf C60} (1993) 83;
 J.~L.~Chkareuli and I.~G.~Gogoladze,
  Phys.\ Rev.\  D {\bf 58}, 055011 (1998).




\bibitem{Pierce:1996zz}
D.~M. Pierce, J.~A. Bagger, K.~T. Matchev, and R.-j. Zhang,
  { Nucl. Phys.} {\bf B491} (1997) 3.




\bibitem{Leva}
J.L. Leva,
 Math. Softw. 18 (1992) 449;
J.L. Leva,
Math. Softw. 18 (1992) 454.




\bibitem{Nakamura:2010zzi}
  J.~Beringer {\it et al.}  [Particle Data Group Collaboration],
  Phys.\ Rev.\ D {\bf 86}, 010001 (2012).



\bibitem{Baer:2002fv}
H.~Baer, C.~Balazs, and A.~Belyaev,
   { JHEP} {\bf 03} (2002) 042;
 H.~Baer, C.~Balazs, J.~Ferrandis, and X.~Tata
  { Phys. Rev.} {\bf D64} (2001)  035004.



\bibitem{:2007kv}
  T.~Aaltonen {\it et al.}  [CDF Collaboration],
  Phys.\ Rev.\ Lett.\  {\bf 100}, 101802 (2008).


\bibitem{Barberio:2008fa}
  E.~Barberio {\it et al.}  [Heavy Flavor Averaging Group],
  arXiv:0808.1297 [hep-ex].










\bibitem{Ajaib:2012vc}
  M.~A.~Ajaib, I.~Gogoladze, F.~Nasir and Q.~Shafi,
  Phys.\ Lett.\ B {\bf 713}, 462 (2012).




\bibitem{Djouadi:2005gj}
  A.~Djouadi,
  Phys.\ Rept.\  {\bf 459}, 1 (2008).



\bibitem{gunion90}
J.~F.~Gunion, H.~E.~Haber, G.~L.~Kane and S.~Dawson,
{\it``The Higgs Hunter's Guide''}, Addison-Wesley, Reading (USA), 1990.
%








\bibitem{Gogoladze:2009bd}
  I.~Gogoladze, M.~U.~Rehman and Q.~Shafi,
  Phys.\ Rev.\ D {\bf 80}, 105002 (2009).












\end{thebibliography}
\end{document}